\documentclass[useAMS,usegraphicx]{mn2e}
\usepackage{times}

\title[SBs and the Field-Binary Population]
{What a Local Sample of Spectroscopic Binaries can tell us
about the Field-Binary Population}

\author[J. Fisher, K.-P. Schr\"oder, Robert Connon Smith]{
        James Fisher\thanks{E-mail: jfisher@sussex.ac.uk},
        Klaus-Peter Schr\"oder,
        Robert Connon Smith\\
        Astronomy Centre, Department of Physics and Astronomy,
        University of Sussex, Falmer, Brighton, BN1 9QH, UK
}

\begin{document}
\date{March 2005: revised version of MN paper ME792}
\pagerange{\pageref{firstpage}--\pageref{lastpage}}
\maketitle
\label{firstpage}


\begin{abstract}

We study a sample of spectroscopic binaries (SBs) in the local solar neighbourhood
($d~\le~100\,$pc and $M_{\rm{V}} \le 4$) in an attempt to find the distributions
of the period, $P$, the primary mass, $m_1$, and the mass ratio $q~(=m_2/m_1)$, as
well as the IMF of the local population of field binaries. The sample was collated
using available SB data and the \emph{Hipparcos} catalogue, the latter being used
for distances and to refer numbers of objects to fractions of the local stellar
population as a whole. We use the better-determined double-lined SBs (SB2s) to
calibrate a Monte-Carlo approach to modelling the $q$ distribution of the
single-lined SBs (SB1s) from their mass functions, $f(m)$, and primary masses,
$m_1$. The \emph{total} $q$ distribution is then found by adding the observed SB2
distribution to the Monte-Carlo SB1 distribution. While a complete sample is not
possible, given the data available, we are able to address important questions of
incompleteness and parameter-specific biases by comparing subsamples of SBs with
different ranges in parameter space. Our results show a clear peak in the $q$
distribution of field binaries near unity. This is dominated by the SB2s, but the
flat distribution of the SB1s is inconsistent with their components being chosen
independently at random from a steep IMF.

\end{abstract}



        \section{Introduction}
        \label{s:intro}


Much has been published on binary statistics and much of this has been concerned
with the statistics of spectroscopic binaries (SBs) (e.g. Goldberg, Mazeh \&
Latham 2003; Halbwachs et al. 2003; Boffin, Cerf \& Paulus 1993; Boffin, Paulus \&
Cerf 1992). Among the statistics discussed three of the most important have been
the distributions of period ($P$), primary mass ($m_1$) and mass-ratio
($q\,=\,m_2/m_1$, where the primary in this paper is always taken to be the
more massive star). These three parameters in particular are important as they
present a set of variables that suffice to describe the principal properties of a
binary system and its evolutionary path. Many papers have also been published
specifically just on the mass-ratio distributions of SBs (e.g. Hogeveen 1991;
Trimble 1974, 1978, 1987, 1990). The question of whether there is a peak in the
$q$ distribution near $q = 1$ has been a matter of debate for some time, on which
we hope to shed some light in this paper. For instance the recent paper of
Goldberg et al. (2003) found a distinct bimodal distribution; however, unlike
ours, their sample was not confined to a particular volume and so might be
expected to exhibit biases related to this fact.

The Initial Mass Function (IMF) of binaries is also of interest and, together with
the period distribution, is important for the validation of star formation models.
The IMF and $P$ distributions are also important for understanding the chemical
evolution of the Galaxy: interacting binary systems have more complex evolutionary
pathways by which material can be lost to the Interstellar Medium (ISM) than
single stars, leading, for example, to important systematic corrections to
predictions of the carbon yield (Tout et al. 1999). Binary-system population
synthesis models may then give a more complete idea of the enrichment of chemical
elements in the ISM. This kind of modelling has been done before for single stars
but to date studies of this kind for double stars have had to make certain
assumptions about the $P$, $m_1$ and $q$ distributions and the IMF.

There have also been studies of stars within galactic clusters (e.g. K\"ahler
1999) which are thus magnitude-limited and occupy specific limited volumes, as is
the case for our sample, but which constitute very specific subsets of stars all
of a particular age. Cluster stars are also in the process of diffusing from the
cluster at a significant rate over time-scales of the order of $10^8$ yrs. Hence
the statistics of cluster binaries, while interesting in their own right, are not
necessarily representative of binaries in general, or field binaries in particular
and we do not consider them further.

What has not been done before to our knowledge is a study of the statistics of a
distance and luminosity-limited sample of field binaries that is, as far as is
possible with existing data, evolutionarily unbiased and complete.

In this paper we derive the distributions for the period $P$, primary mass $m_1$,
mass ratio $q = m_2/m_1$ ($m_2$ being the mass of the secondary so that $q<1$) and
the IMF for a distance and luminosity-limited sample of SBs in the local solar
neighbourhood $d \le 100$\,pc and $M_{\rm{V}} \le 4$ and from there are able to
make some deductions about these same distributions for the general population of
field binaries. Because our sample is luminosity-limited, the only systems we will
be missing completely will be those with primary mass $m_1 < 1.1 \, 
\mathrm{M_\odot}$ on the Main Sequence.
This luminosity cut-off makes the sample as unbiased as possible in an
evolutionary sense.

While the sample is certainly \emph{not} complete to the distance and luminosity
limits stated, we believe that it is as complete as it is possible to make it
without seriously compromising the sample size, and so without also compromising
the conclusions drawn from the study. Although the sample is far from complete in
absolute terms, we are able to make an assessment of the incompletenesses and are
thus able to attempt to compensate for them. We thus believe that our sample is
the best approximation to a volume-limited sample possible with the data currently
available (to be truly volume-limited the sample would have to include \emph{all}
objects to the specified distance and luminosity limits).


        \section{The Sample}


The spectroscopic binary data for the study was taken from the Eighth Catalogue of
Orbital Elements of Spectroscopic Binary Systems (Batten, Fletcher \& MacCarthy
1989), hereafter referred to in this paper as the `Batten' catalogue, supplemented
by other data of R.F. Griffin, both published (see the synopsis paper Griffin 2000
and others of that series) and unpublished (private communication), hereafter
referred to as `RFG' data. The Batten catalogue was until recently\footnote{The
Ninth Catalogue has just appeared (Pourbaix et al. 2004).} the most
comprehensive catalogue of SBs available, its selection criterion simply being to
include all SB data available at the time of compilation. It thus encompasses a
very diverse range of motivations for observation from all the many contributors,
and so will contain a variety of selection effects, which are largely unknown
apart from the tendency mentioned below to favour shorter-period systems. The
inclusion of the RFG data was designed firstly to increase the size of the
available dataset (many more objects having been observed since publication of the
Batten catalogue in 1989, many of them by Roger Griffin), and secondly to attempt
to compensate for the inevitable bias of the Batten catalogue towards
shorter-period systems (inevitable because of the difficulties involved in
sustaining consistent observing programmes for longer-period systems, as Roger
Griffin has been able to do).

The Batten catalogue consists of 1469 SBs and the RFG data of 498 SBs. The
combined Batten and RFG dataset (after removing duplicated objects) was filtered
to a distance of 100\,pc and limiting absolute magnitude of 4 ($d \le 100$\,pc and
$M_{\rm{V}} \le 4$) by correlating entries with the \emph{Hipparcos} catalogue.
The correlation was done by Henry Draper catalogue (HD) number, or by coordinates (corrected for
precession) if no HD number existed, using the \emph{Hipparcos} parallaxes and
apparent magnitudes to calculate distances and absolute magnitudes. The cutoff for
absolute magnitude was chosen in such a way that the sample should (to that
magnitude) be as complete and homogeneous as possible, while at the same time
having a sufficient number of systems to derive reasonable statistics. The chosen
limiting absolute magnitude of 4.0 translates to an apparent magnitude of 7.5 at
50\,pc and to 9.0 at 100\,pc. The \emph{Hipparcos} catalogue is complete to
$m_{\rm V}= 7.3$, and in some areas down to $m_{\rm V} = 9.0$ (Perryman et~al.
1997; Schr\"oder \& Pagel 2003), so all the SBs in our initial dataset of 1803
distinct stars that are closer than 50\,pc (more precisely 46\,pc) and brighter
than $M_{\rm{V}} = 4$ should have been identified via the comparison with
\emph{Hipparcos}, and a reasonable fraction of those closer than 100\,pc should
also have been identified. A fainter limit would have created an inconsistent
sample that, while having data from more systems at close distances, would be
missing a lot of fainter systems at greater distances; it is already clear from
the number in our final sample that many of the systems in the SB catalogues lie
beyond 100 pc. The fainter limit would have led to a much steeper fall-off of
completeness with distance than we find in our chosen sample, creating a sample
that was harder to analyse. As mentioned above, because our sample is
luminosity-limited with an absolute magnitude limit of ($M_{\rm{V}} \le 4$), the
only systems we will be missing completely will be those with primary mass $m_1 <
1.1 \, \mathrm{M_\odot}$ on the Main Sequence.

When filtered in this way, the sample consists of 371 SBs: 145 double-lined SBs
(SB2s) and 226 single-lined SBs (SB1s). It is this sample that we work with in the
analysis that follows in the rest of the paper.


        \section{Distributions}

        \subsection{Period ($P$) distribution}

The period distribution was found directly from the spectroscopic
binary data. For later use in evolutionary studies the periods were
divided into the four categories given in
Table~\ref{table:period/evol-cats}. It is intended that the four
period categories should correspond roughly to the following four
evolutionary scenarios:

\begin{enumerate}
        \item $P\ge 500$ d. \ Systems that will
         interact, if at all, at a late stage in their evolution,
     with the primary well on its way up the Asymptotic Giant Branch (AGB).
    \item 500 d $>P\ge$ 10 d. \ Systems that
         interact as the primary evolves onto the Red Giant
         Branch (RGB) or the AGB.
    \item 10 d $>P\ge$ 1 d. \ Systems that interact on the early
         RGB or as the primary leaves the Main-Sequence and enters the
         `Hertzsprung Gap'.
    \item $P<1$ d. \ Systems that interact while on
         the late Main Sequence or earlier (contact systems).
\end{enumerate}

\begin{table}
        \caption{Period/evolutionary categories. The numbers of SBs
        are for $d \le 100$\,pc and  $M_{\rm{V}} \le 4$.}

    \begin{tabular}{ccrrr}
        \hline
    Category &       Period range\,/\,days   & SB1s & SB2s & Total SBs\\
        \hline
    (i)      &       $\;\;\,P\ge$ 500    &   73 &   15 &        88\\
    (ii)     &  500  $>P\ge$       10    &   84 &   60 &       144\\
    (iii)    &   10  $>P\ge$        1    &   62 &   63 &       125\\
    (iv)     &       $\;\;P\,<  $   1    &    7 &    7 &        14\\
    \hline
    TOTALS   &                           &  226 &  145 &       371\\
    \hline

        \end{tabular} \label{table:period/evol-cats}
\end{table}

\begin{figure}
    \includegraphics[scale=0.35,angle=-90]{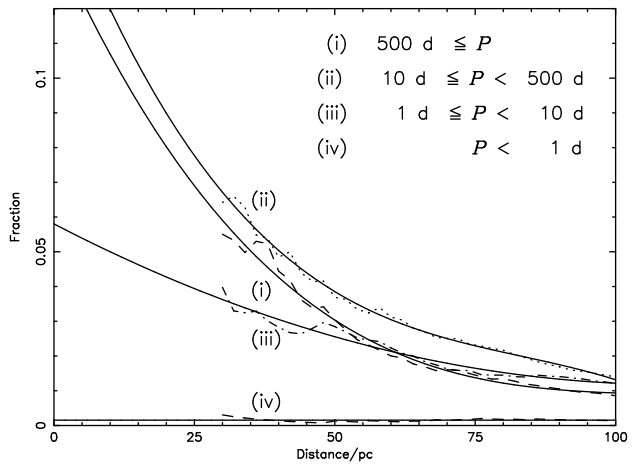} 
    \caption{Period distributions of all SBs in the sample for the four period categories
    given in Table~\ref{table:period/evol-cats}, together with
    extrapolations to 0\,pc. Fractions to a particular distance are the fractions
    of \emph{all} systems (single and multiple) with $M_{\rm V} \le 4$ in the
    \emph{Hipparcos} catalogue to that distance.}  \label{fig:Pdistribs}
\end{figure}

To investigate the behaviour of the observed period distributions within different
volumes \emph{within} our 100\,pc sample, the fractions of SBs (of all systems,
single and multiple, determined from \emph{Hipparcos}) in the above period
categories were determined for volumes of radius 20 to 100\,pc in steps of 2\,pc
for $M_{\rm{V}} \le 4$. The absolute magnitude cutoff (the same as for the rest of
the study) avoids the increasing incompleteness that would otherwise result with
increasing distance (without the cutoff we would include too many faint stars,
many of which could reasonably be supposed to be unrecognised binaries). The
results are shown in Fig.~\ref{fig:Pdistribs}. The figure demonstrates well the
quality of the data to 30\,pc. For volumes much smaller than $d = 30$\,pc the data
are of little statistical significance due to the low absolute numbers involved
(e.g. at 25\,pc there are 12, 11, 8 and 0 systems in $P$ categories  (i) to (iv)
respectively out of a total of 284 \emph{Hipparcos} objects). The low numbers cause
the fractions to vary somewhat erratically below 30\,pc and so are also of little
use in determining the trends. The results will be discussed in more detail in
Section~\ref{sect:discussion}.

Clearly, a significant fraction of the systems are missing, but we can estimate
the complete fractions for each category by extrapolation to 0\,pc. Although the
data are for SBs, the extrapolations will necessarily be the fractions of
\emph{all} binary/multiples (with respect to all stellar objects, single and
multiple) as all binaries will be detectable as SBs at 0\,pc. The extrapolation
works because the further out in the sample we go, the less likely it is that a
binary system is known as an SB. Therefore, as we go in the opposite direction,
towards 0\,pc, the fraction of systems that are detectable as SBs increases, and
we get closer to the values of the unbiassed fractions. Fitting curves to the data
from 100\,pc to 30\,pc, we were thus able to determine what the complete fractions
\emph{would} be at $d=0$\,pc.

Categories (i) and (ii) were fitted by cubic polynomials and category (iii) by a
quadratic curve. The choice of curve was determined by its shape as well as the
closeness of the fit: quadratics for categories (i) and (ii) and a cubic for
category (iii) all gave curves that changed from convex to concave with distance
and hence were not plausible. A cubic fit may be preferred on the grounds that it
is what might be expected for number density within a spherical volume (the
100\,pc limit is not large enough for the thickness of the galactic disc to become
relevant (Schr\"oder \& Pagel 2003), in which case a quadratic fit might have been
expected instead) but they are also more suseptible to errors in extrapolations
due to the higher order term. The levelling out of the curves at larger distances
is roughly what one would expect on the basis of the kind of bias expected at
different distance regimes. At closer distances an apparent magnitude bias would
be expected, as the brighter a system's $m_{\rm V}$ the more likely it would be to
be known as an SB, while at larger distances one would expect a random selection
bias due to the effectively random basis on which systems are chosen for study,
causing the fraction of SBs to tend to a constant value at larger distances. For
category (iv) it wasn't possible, or realistic, to do anything more with the data
than to make a linear fit over the same distance range, this category being
considerably limited in what we can do with it by its low absolute numbers.

The fits were made over as large a range of distances as possible before low
absolute numbers made the data unreliable, in each case from 100\,pc to 30\,pc.
Clearly it is difficult to estimate the errors on such extrapolations and the
fractions are not necessarily going to be very accurate. However the fractions
given by quadratic fits for categories (i) and (ii) instead of the cubic fits used
can be used to give some idea of the errors involved. In both case the quadratic
fits had lower constants, $D$, and hence lower fractions at 0\,pc. Taking these
differences into account the accuracy of the extrapolations is estimated to be of
the order of 10\%. Details for all four categories are given in
Table~\ref{table:period_polyfits}. The extrapolations give the following estimates
for the complete fractions of SBs in the four period categories (equal to the
value of the constants $D$ in Table~\ref{table:period_polyfits}): 0.140, 0.158,
0.0580 and 0.00154, giving a total binary fraction of $0.36 \pm 0.11$ (for stars
brighter than $M_{\rm V} = 4$).

\begin{table}
        \caption{Polynomial fits for Period categories:  {\em
        Fraction} $= Ad^3+Bd^2+Cd+D$, where $d$ is distance in pc.
    There are 36 data for each
        category (30\,pc to 100\,pc in steps of 2\,pc).
    The constant $D$ is also equal to the
        extrapolated fraction at 0\,pc.}

    \begin{tabular}{ccccc}

        \hline
               & $A$                   & $B$                  & $C$
               & $D 
               $               \\
        \hline
      (i)  & $-1.05 \cdot 10^{-7}$ & $3.33 \cdot 10^{-5}$ & $-3.59 \cdot 10^{-3}$  & $0.140$             \\
     (ii)  & $-1.89 \cdot 10^{-7}$ & $4.72 \cdot 10^{-5}$ & $-4.28 \cdot 10^{-3}$  & $0.158$             \\
    (iii)  & $0$                   & $3.82 \cdot 10^{-6}$ & $-8.40 \cdot 10^{-4}$  & $5.80 \cdot 10^{-2}$\\
         (iv)  & $0$                   & $0$                  & $0$                    & $1.54 \cdot 10^{-3}$\\
        \hline
        \end{tabular}
        \label{table:period_polyfits}
\end{table}

The above completes the more general analysis of the incompleteness of
the sample. Selective incompletenesses will be considered in
Section~\ref{sect:discussion}.

        \subsection{Primary-mass ($m_1$) distribution}
        \label{subsect:Primary-mass distrib}

Firstly, the \emph{Hipparcos} parallaxes and apparent visual
magnitudes of the systems, $m_{\rm{V}}$, were used to calculate the
absolute visual magnitudes, $M_{\rm{V}}$. The primary masses of the
SBs were then estimated from $M_{\rm{V}}$ by correcting for an average
contribution of the secondary and applying a mass-luminosity
relationship.

The absolute magnitude of the primaries, $M_{\rm{V}_1}$, were estimated
using the following magnitude offsets from the absolute magnitude of
the system, $M_{\rm{V}}$:
\begin{eqnarray}
        M_{\rm{V}_1} = M_{\rm{V}} + 0.50 \mathrm{\ \ (for\ SB2s)}\\
    M_{\rm{V}_1} = M_{\rm{V}} + 0.20 \mathrm{\ \ (for\ SB1s)}\label{eqn:SB1_offset}
\end{eqnarray}
\\
The offsets are necessary as the presence of even a visually unseen
companion can make a significant difference to the magnitude of the system
(see the discussion of $\zeta$ Aurigae systems below).

The offset for SB2s was determined by the clear peak of their $q$ values
(Section~\ref{subsect:q distrib}) near 1, with an average $q$ of $\approx 0.84$
(see Fig.~\ref{fig:qSB2and1} in that section). This means that the average SB2
secondary is about 40 per cent less luminous than the primary and so would
contribute about 0.5 mag to the system's $M_{\rm{V}}$. (It should be noted in
passing that the individual magnitudes of the SB primaries and secondaries are not
usually known, even in the case of SB2s. Where the Batten catalogue sometimes
quotes two magnitudes these are in fact the maximum and minimum magnitudes of the
system if the apparent magnitude of the system is variable.)

The offset for SB1s is a mean figure suggested by two considerations. Firstly, if
the luminosity of the secondary were, on average, 30 per cent less than that of
the primary (equivalent to a magnitude difference between the primary and the
system of 0.3) then the contrast of the secondary's spectral lines would be enough
for them to be visible, and the system would be observed as an SB2 rather than an
SB1. Secondly, the depths of eclipse in $\zeta$ Aurigae systems (typically also
catalogued as SBs, for example $\zeta$ Aurigae itself is in the Batten catalogue)
are equivalent to the secondary contributing 0.1 - 0.2 mag. to the magnitude of
the system. (See, for example, the photometry of the January 1989 eclipse of
$\tau$ Persei given in Hall et al. 1991 and of the 1988 eclipse of 22 Vulpeculae
in Griffin et al. 1993. It should also be noted that the only reason that $\zeta$
Aurigae and 22 Vulpeculae are known as SB2s is because their secondaries are
visible in the ultraviolet band. For observations of $\zeta$ Aurigae systems in
the visible waveband they are usually only seen as SB1s, as for example is 22 Vul
which is given in the Batten catalogue as an SB1.) From these two considerations
we therefore adopted an offset of 0.2 mag to account for an average SB1
secondary's contribution to the luminosity of the system.

Alternatively, we can argue from the $q$ distribution that results from our
calculations. From Fig.~\ref{fig:qSB2and1}, we see that typically $q\simeq0.5$,
with a rather large uncertainty because of the flat distribution.
Using the same argument as above for SB2s, this would lead to a
contribution of about 0.13 to $M_{\rm V}$, which is (within the uncertainties)
consistent with the 0.2 offset we have assumed. To test the effect of the
offset, we have run the calculation below again with a zero offset, and the
results are qualitatively similar (see Section~\ref{subsect:q distrib}).

\begin{figure}
    \includegraphics[scale=0.45]{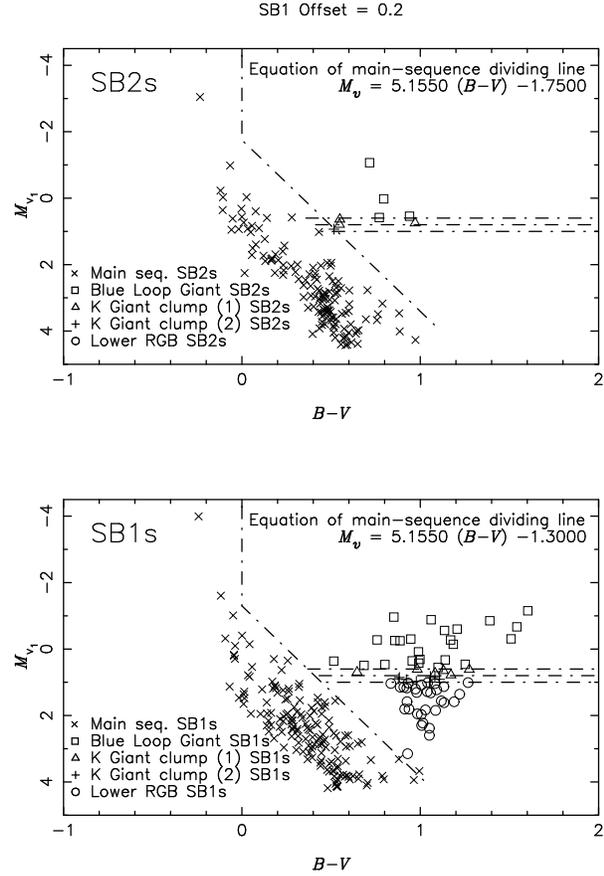} 
    \caption{Evolutionary categories for SB2s and SB1s. These are used to
determine which mass-luminosity relationship to use in estimating the masses
of the primaries.}
    \label{fig:evolHRD}
\end{figure}

A number of different mass-luminosity relationships were used to
 determine $m_1$ from $M_{\rm{V}_{1}}$ according to the evolutionary
 status of the primary. The evolutionary status was determined from its location on an
 Hertzsprung-Russell Diagram (HRD) using the $B\!\!-\!\!V$ colour index from the \emph{Hipparcos}
 catalogue and the value of $M_{\rm{V}_1}$ already calculated. The HRD
 was divided into a number of regions based
 on the characteristic regions used by Schr\"oder
 (1998) and Schr\"oder \& Sedlmayr (2001) (see
 Fig.~\ref{fig:evolHRD}).  For main sequence stars the mass-luminosity
 relationship used was for stars half-way through their H
 core-burning phase obtained from detailed theoretical stellar models
 computed with the well-tested evolution code of Peter Eggleton (Pols
 et al. 1998). The other mass-luminosity relationships are from
 Schr\"oder \& Sedlmayr (2001) and Schr\"oder (1998). We thus have the
 following set of equations for $m_1$: \\
\\For Main Sequence primaries, both
\begin{eqnarray}
        B\!-\!V & < & (M_{\rm{V}_1} + 1.50)/5.16\\
        \&\quad\quad M_{\rm{V}_1} & \geq & -1.50
\end{eqnarray}
and
\begin{eqnarray}
        B\!-\!V & < & 0\\
        \&\quad\quad M_{\rm{V}_1} & < & -1.50
\end{eqnarray}
use the following mass-luminosity relation
\\
\begin{eqnarray}
\nonumber 
        m_1 & = & (3.57 - 1.40M_{\rm{V}_1} + 0.311M_{\rm{V}_1}{}^2\\
            &   & \ \ - 0.027M_{\rm{V}_1}{}^3)\ \mathrm{M_\odot}
\label{eqn:mass_main-seq}
\end{eqnarray}
For Blue-loop giants:
\begin{eqnarray}
        B\!-\!V & \geq & (M_{\rm{V}_1} + 1.50)/5.16\\
        M_{\rm{V}_1} & < & 0.6\\
        m_1 & = & (-0.852M_{\rm{V}_1} + 2.81)\ \mathrm{M_\odot}
\label{eqn:mass_BL}
\end{eqnarray}
For K giant clump stars (subgroup 1):
\begin{eqnarray}
        B\!-\!V & \geq & (M_{\rm{V}_1} + 1.50)/5.16\\
        0.6\ \ \le & M_{\rm{V}_1} & <\ \ 0.8\\
        m_1 & = & 1.8\ \mathrm{M_\odot} \label{eqn:mass_K1}
\end{eqnarray}
For K giant clump stars (subgroup 2):
\begin{eqnarray}
        B\!-\!V & \geq & (M_{\rm{V}_1} + 1.50)/5.16\\
        0.8\ \ \le & M_{\rm{V}_1} & <\ \ 1.0\\
        m_1 & = & (-0.852M_{\rm{V}_1} + 2.2)\ \mathrm{M_\odot} \label{eqn:mass_K2}
\end{eqnarray}
For Lower RGB giants:
\begin{eqnarray}
        B\!-\!V & \geq & (M_{\rm{V}_1} + 1.50)/5.16\\
        M_{\rm{V}_1} & \ge & 1.0\\
        m_1 & = & 1.25\ \mathrm{M_\odot} \label{eqn:mass_lower_RGB}
\end{eqnarray}

\begin{figure}
    \includegraphics[scale=0.50]{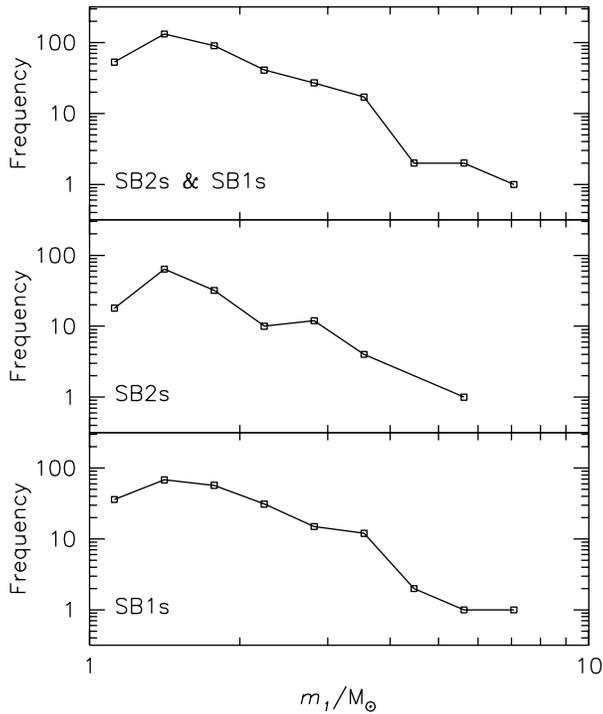} 
    \caption{$m_1$ distributions determined in Section~\ref{subsect:Primary-mass
distrib}.}
    \label{fig:m1}
\end{figure}

The resulting distributions of estimated primary masses are
shown in Fig.~\ref{fig:m1}.

        \subsection{Mass-ratio ($q = m_2/m_1$) distribution}
        \label{subsect:q distrib}

The $q$ distribution for SB2s was found directly from the observed
orbital semi-amplitudes, $K_2$ and $K_1$, the period, $P$, and the
eccentricity, $e$, via:

\begin{equation}
      \label{eqn:msin3i}
      m_{1,2} \sin^3 i = 1.036\cdot10^{-7} (1-e^2)^{3/2} (K_2 + K_1)^2 K_{2,1} P\\
\end{equation}
where $m_{1,2}$ have units of solar masses (M$_\odot$), $K_{1,2}$ have
units of km\,s$^{-1}$ and $P$ has units of days (Hilditch 2001, p46). Hence:

\begin{equation}
        q = \frac{m_2}{m_1} = \frac{m_2 \sin^3 i}{m_1 \sin^3 i} = \frac{K_1}{K_2}.
\end{equation}

The $q$ distribution for SB1s however is not so easy to determine.
The closest one can get to $q$ directly is the following function of
the mass function, $f(m)$, and the primary mass, $m_1$:

\begin{equation}
    \label{eqn:f(m)/m1}
    \frac{f(m)}{m_1} = \frac{q^3 \sin^3 i}{(1 + q)^2}.
\end{equation}
The mass function is calculated from the observed period, $P$,
primary orbital semi-amplitude, $K_1$, and eccentricity, $e$,
\begin{equation}
    \label{eqn:fm}
    f(m) = 1.036\cdot 10^{-7} \,(1-e^2)^{3/2} \,{K_1}^3 \,P
\end{equation}
using the same units as for equation~\ref{eqn:msin3i}. The primary mass, $m_1$,
may be determined as in Section~\ref{subsect:Primary-mass distrib}.

There are a number of different methods for determining the underlying
$q$ distribution from the $f(m)/m_1$ distribution. Two of the most
commonly used are:

\begin{enumerate}
        \renewcommand{\theenumi}{(\arabic{enumi})}

        \item  Richardson-Lucy iterative method (not used in this paper). This is a method that
was first developed by Richardson (1972) for image restoration in
optics and
then first adapted for astronomical use by Lucy (1974) for the deconvolution
of unknown distributions. It has since been used on a number of occasions for
deconvolving $q$ distributions from observed distributions
(e.g. Hogeveen 1991). As it is not used in this paper no further
details of this method shall be given here.
\\
        \item  Monte-Carlo simulation (used in this paper in a refined form). This involves calculating the $q^3\sin^3 i/(1 + q)^2$
distributions from a variety of postulated $q$ distributions and matching
them to the observed $q^3\sin^3 i/(1 + q)^2$ distribution.

\end{enumerate}
[For other methods see, for instance, Halbwachs (1987).]
\\

While the Monte-Carlo method might be considered to be
somewhat unsophisticated, especially given some of the
assumptions that have had to be made in the past to make the method
work (see below), we have been able to introduce a number of
constraints and tests that we hope improve it somewhat and make it
more robust. By contrast, methods such as Richardson-Lucy are not as
direct and are also more dependent on initial assumptions.

        \subsubsection{A short review of previous Monte-Carlo studies}

This method has been used on many occasions before, but a number of
 restrictions and/or assumptions have always had to be made in order
 to make it work. For example Boffin et al. (1993) restricted
 the sample to 213 spectroscopic binaries with red giant primaries,
 assumed a constant mass for the primaries and also an average value of
 $\sin^3 i$. Trimble's study (Trimble 1990) was also limited to a
 subset of SBs, 164 in this case, most of which were K giants, their
 primary  masses being determined from their spectral classes.

A major problem encountered in Monte-Carlo determinations of $q$
distributions is how to take  account of the unknown orbital
inclination angle $i$. The effect of any assumed $i$ distribution is
made more critical by the dependence of  $f(m)/m_1$ on $\sin^3
i$. Typically the unknown $i$ s have been accommodated  by assuming an
average value for $\sin^3 i$. As well as being a bit crude, and so
perhaps of better use where more rigorous methods may not be possible,
this method does have other distinct problems. Boffin et al. (1993)
show that, using this method, it is possible to obtain correct looking
results for a decreasing $q$ distribution while giving a totally wrong
result for the case where $f(q) \propto 1/q$. In fact, for this reason
Boffin et al. (1993) discard a simple Monte-Carlo approach despite
others such as Trimble (1990) and Hogeveen (1991) finding that it
gives results similar to more sophisticated approaches such as the
Richardson-Lucy method. Mazeh \& Goldberg (1992) also find that it
produces erroneous results; in their paper they show two graphs of
simulations with invented $q$ distributions and demonstrate how badly they
are reconstructed using this method: an even distribution is reconstructed
as a decreasing distribution with constant gradient, and a distribution
increasing towards $q=1$ is reconstructed as an upside-down U-shaped
distribution. They then go on to show that these results are a
consequence of some of the initial assumptions of the method being
invalid. Another procedure is to adopt
a `model-fitting' approach where the probability of detecting a system
with a certain inclination, $i$, is set by theoretical
considerations. This however usually involves making somewhat \emph{ad
hoc} theoretical assumptions.

Trimble (1974, 1990) took two approaches to the unknown $\sin^3 i$ values: the
direct approach using an average value of $\sin^3 i$, and a model-fitting approach
that assumed the probability, $p$, of there being an orbit of inclination, $i$, to
be proportional to a certain function: $p(i) \propto \sin i$ in Trimble (1990), as
suggested in Halbwachs (1987), and $p(i) \propto \sin^2 i$ in Trimble (1974).
However the average $\sin^3 i$ values were themselves determined by assuming that
the probability of detecting a system was proportional to a particular function.
Halbwachs (1987) used three methods: the average $\sin^3 i$ method, and two other
methods not discussed here: one due to Abt \& Levy (1985) and the other due to
Jaschek \& Ferrer (1972). None of these methods is entirely satisfactory.

        \subsubsection{Refined Monte-Carlo Method (used in this
      paper)}
    \label{subsubsect:refined_monte_carlo}

In the present study we attempt to avoid some of the problems previously
encountered with this method by introducing some of our own refinements, ones that
are largely made possible by the use of the \emph{Hipparcos} catalogue. For
instance, the subset of better-known SB2s was used to derive constraints that
could then be applied to all SBs. For the masses, we were able to go a step
further than previous studies by using the \emph{Hipparcos} parallaxes to
determine absolute magnitudes, and from there obtain the primary masses directly
from their evolutionary status determined from their position on an HRD (as
discussed above in section~\ref{subsect:Primary-mass distrib}). We thus
circumvented the dual problems of having to make assumptions about the unknown
masses and of being restricted to a limited sample of stars of a particular
luminosity class. The study was hence opened up to stars of any evolutionary
status, subject only to a limiting absolute magnitude of 4 as in the rest of our
study.

We also used a method of random inclinations, $i$, to avoid having to assume an
average value of $\sin^3 i$. The angle $i$ varies between $0^\circ$ and
$180^\circ$, but in practice systems with inclinations near $0^\circ$ or
$180^\circ$ will not be easily visible as the components of their radial velocity
in our line of sight will be too small to measure reliably. Systems with $i$ near
to $0^\circ$ or $180^\circ$ will thus tend to be missed in surveys. For our
Monte-Carlo procedure we have assumed therefore that $i$ varies between a minimum
cutoff angle $\alpha_0$ and $90^\circ$ (sufficient as $i$ only appears as $\sin^3
i$ in the equations). The probability of detecting a system is therefore zero for
$i$ less than $\alpha_0$ and proportional to $\sin i$ between $\alpha_0$ and
90$^\circ$, i.e. $p(i) = 0$ for $i < \alpha_0$ and $p(i) \propto \sin i$ for
$\alpha_0 \le i \le 90^\circ$. The reason for the proportionality to $\sin i$ is
that the projection of the SB's orbit onto the line of sight is proportional to
$\sin i$, and hence so is the probability of observing the system as an SB. This
means that a initial uniformly random variate, $x$, has to be transformed to $i =
\arccos (1 - x)$ (the mathematical reasoning behind the transformation is given in
section 7.2 of Press et al. 1993). This is, we think, more reasonable than
previous assumptions that have been made (and indeed have had to be made).

A problem with the estimated mass procedure as implemented is that the
masses for the main sequence primaries (that is, the greater
proportion of them) are taken from half-way through their H
core-burning phase. In reality the masses will be `smudged out' to
either side of these values. To simulate this when calculating the
$f(m)/m_1$ distribution from random inclinations a `Smudge Factor',
$\zeta$, was therefore introduced so that the simulated $f(m)/m_1$ values are
multiplied by a random factor of up to $(1\pm\zeta)$.

\begin{figure}
    \includegraphics[scale=0.9]{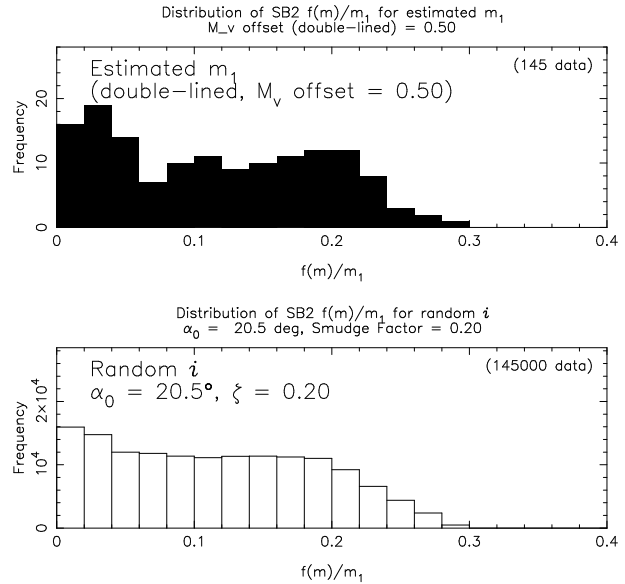} 
    \caption{Examples of histograms for the $\alpha_0$ calibration
    taken from the two sources given in
    Section~\ref{subsubsect:refined_monte_carlo}. The $\zeta$
    calibration has already been performed so that the maximum values of $f(m)/m_1$
    are the same from both sources.}
    \label{fig:alpha_0_cal}
\end{figure}

The values of $\alpha_0$ and $\zeta$ were found by comparing the $f(m)/m_1$
distribution for SB2s from two different sources: (i) from their $q$
values and random $i$ values, $\alpha_0 \le i \leq 90^\circ$ for
different values of $\alpha_0$ and $\zeta$ (1000 random $i$  values per $q$ value) and (ii)
from their  mass functions, $f(m)$, and estimated primary masses,
$m_1$. The $f(m)/m_1$ distributions were plotted as histograms, 5 bins
per 0.1 on the $f(m)/m_1$ axis (see Fig.~\ref{fig:alpha_0_cal} for an
example of the histograms from each source for the $\alpha_0$ calibration).

The main effect of the smudging out of the masses on the $f(m)/m_1$
distribution from (ii) was that the maximum value of $f(m)/m_1$ became
greater than the maximum theoretical value  of 0.25. (A smaller
additional effect was that the slope of the distribution towards the
maximum $f(m)/m_1$ value was slightly shallower than that from (i).)
From a range of values of $\zeta$ from 0.0 to 0.5 in steps of 0.1,
$\zeta = 0.20 \pm 0.05$ was selected as giving the same maximum value
of $q^3\sin^3 i/(1 + q^2) = f(m)/m_1$, from (i) as from (ii).

Once $\zeta$ had been determined in this way the value of $\alpha_0$
could then be found by comparing the ratios of the first bin
($f(m)/m_1$ from 0 to 0.025) to the total frequency. By this means the
best match was found to be for $\alpha_0 = 20.5\pm1.0^\circ$.

The determination of $\alpha_0$ and $\zeta$ were both cases demonstrating how the
better known SB2s could be used to calibrate parameters for the Monte-Carlo
approach to the SB1s. Not only were we able to validate methods before using them
for the unknown SB1 $q$ distribution but we were able to determine fine-tuning
parameters for the simulation as a whole. We did then have to assume that the
values of $\alpha_0$ and $\zeta$ were the same for SB1s as for SB2s. This would be
reasonable if the SB2s and SB1s have similar selection criteria and if the
probabilities of detecting a spectroscopic binary as an SB2 or an SB1 are
independent. The similar behaviour of the fractions of SB2s and SB1s in
Fig.~\ref{fig:fracs_M_v} in Section~\ref{sect:discussion} gives us some confidence
that this is indeed the case.

To perform the Monte-Carlo simulation a variety of plausible $q$ distributions
were constructed by dividing the range of $q$ values from 0 to 1 into ten equal
bins and choosing frequencies for each bin. The resultant $f(m)/m_1$ distributions
were then calculated for random values of the inclination, $i$, from $\alpha_0$ to
$90^\circ$, $p(i) \propto \sin i$, with random values of the smudge factor between
$1-\zeta$ and $1+\zeta$. For each data point on the $q$ distribution a thousand
random values of $i$ were used to make the resultant $f(m)/m_1$ distribution as
smooth as possible. The total frequencies for the constructed $f(m)/m_1$
distributions ranged from $2\times10^4$ to $2\times10^5$. The $f(m)/m_1$
distribution was divided in each case into 25 bins from 0 to 0.25 (the maximum
value of $f(m)/m_1$ for $q\le1$ being 0.25). We then had to assume, albeit from
reasonable arguments, the offset to add to $M_{\rm{V}}$ to determine
$M_{\rm{V}_1}$ for the SB1s (see equation~\ref{eqn:SB1_offset} and the text
afterwards).

The $q$ distributions tried were systematic variations on the following `types':
exponential-like functions increasing in frequency towards $q=1$ but with the rise
starting at varying $q$ values, `hump-functions' with a pronounced maximum at
varying $q$ values and different levels upon either side, and `step-functions'
with the step at varying $q$ values and of varying size. As the resultant
$f(m)/m_1$ distributions proved to be rather insensitive to many of the variations
tried we settled on a step function as being a minimal solution requiring the
fewest arbitrary assumptions.

However, given that other authors have found a preference for a bimodal
distribution of $q$, with a secondary peak at around $q=0.2$ (e.g. Staniucha 1979), 
we also present
results from peaked $q$ distributions to see whether the resulting $f(m)/m_1$
distributions match observations better or worse than the simple step function.

The test used to determine the optimal $q$ distribution was to compare two ratios:
the ratio of the frequency of the first bin ($0<f(m)/m_1\le0.01$) to the second
($0.01<f(m)/m_1\le0.02$) and the ratio of the frequency of the first bin to the
total for the 5th to 25th bins ($0.04<f(m)/m_1\le0.25$). The first gives a measure
of the rise in frequency at the low end of the distribution while the second ratio
summarises the relationship of the size of the peak (invariably near $f(m)/m_1 =
0$) to the rest of the distribution.  The observed ratios for the SB1 $f(m)/m_1$
distribution were 3.250 and 1.492 respectively. The best fit to these ratios for
the $q$ distributions tried was for a $q$ distribution consisting of a level
portion at a value of 20 for $0<q\le0.8$ followed  by a step down to zero and then
continuing at zero for $0.8<q\le1$. When normalised to the total number of
observed SB1s this gave frequencies of 28.3, 28.3, 28.3, 28.3, 28.3, 28.3, 28.3,
28.3, 0 and 0 for the ten equally spaced bins from $q = 0$ to 1.

\begin{figure}
    \includegraphics[scale=0.5]{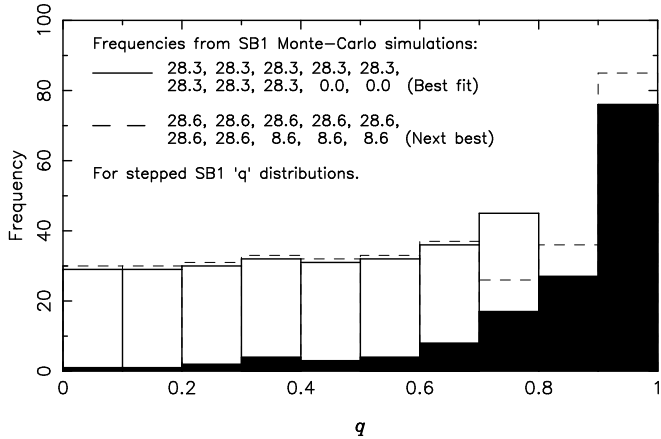} 
    \caption{Histogram of the $q$ distribution for SB2s and SB1s combined, SB2s on
    the bottom (filled-in) and SB1s on top (unfilled), for Monte-Carlo simulations of
    \emph{stepped} SB1 $q$ distributions. The solid line of the SB1s is the best fit,
    the broken line is the next best fit.}
    \label{fig:qSB2and1}
\end{figure}

\begin{figure}
    \includegraphics[scale=0.5]{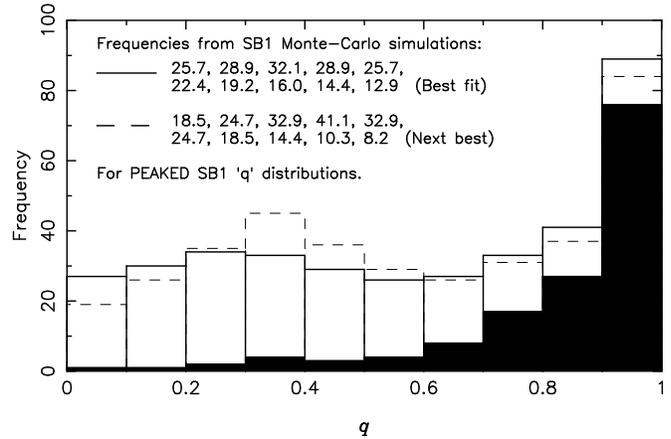}
    \caption{Histogram of $q$ distribution for SB2s and SB1s combined for
    Monte-Carlo simulations of \emph{peaked} SB1 $q$ distributions. SB2s are on
    the bottom (filled-in) and SB1s on top (unfilled). The solid
    line of the SB1s is the best fit, the broken line is the next best fit.}
    \label{fig:qSB2and1-pks}
\end{figure}


\begin{table}
        \caption{A table of the two different sources used in the
      paper for the histograms of combined SB2 and SB1
      $q$ distributions. Given are the (normalised) frequencies for the 10 bins,
      $q = $0 to 1, and a comparison of the main features of the distributions.}
        \begin{tabular}{|l|c|l|}
        \hline
    \bf{Source}  &  \bf{Combined $q$}  &  \bf{General features}\\
    & \bf{distribution} & \\
        \hline
    Observed SB2  & 29, 29, 30, 32, 31, &  Peak, $q = $ 0.9--1.0, \\
    + stepped SB1 & 32, 36, 45, 27, 76  & Plateau, $q = $ 0--0.6,\\
    Monte-Carlo & & at 0.4 of max.\\
        \hline
    Observed SB2  & 27, 30, 34, 33, 29, &  Main peak, $q = $ 0.9--1.0, \\
    + peaked SB1 & 26, 27, 33, 41, 89 & Broad peak, $q = $ 0.2--0.4,\\
    Monte-Carlo & & at 0.38 of max.\\
        \hline
    \end{tabular}
        \label{table:comparison}
\end{table}

The Monte Carlo SB1 distribution is normalised to the total number of SB1s observed
in our sample, to make it directly comparable with the observed SB2 distribution.
If we then simply add the normalised SB1 and observed SB2 distributions with equal 
weight to produce an
overall distribution we are assuming that there is no obvious bias in our sample
towards observing SB2s. Looking at the statistics of SB2s and SB1s in the Batten
catalogue and in the data from Griffin, there is if anything a bias towards SB1s:
our final sample contains 61\%\ SB1s and 39\%\ SB2s. This lack of bias can
probably be explained by the fact that most of the data requires long-term observing
programmes that have only been possible where telescopes have been dedicated to
the programme and the choice of which stars to observe has not been limited by the
exigencies of time allocation committees. Target lists are compiled on the basis
of detected variability, usually long before it is known whether the target is an
SB1 or an SB2, and objects are kept on the target list until an orbit has been
determined. This is
certainly true for the Griffin data and seems likely to be true for the Batten
catalogue as well, most of which dates from the era of long-term programmes at
national or private university telescopes.

Adding the Monte-Carlo SB1 $q$ distribution to the observed SB2 distribution
according to this equal weight prescription gives the combined SB $q$ distribution
shown for the step function in Fig.~\ref{fig:qSB2and1}.  This figure also shows
the effect of adding the next best-fitting SB1 $q$ distribution, the difference
being slight as far as the overall shape of the total distribution is concerned.
The figure clearly shows a peak towards $q=1$. Furthermore, the peak comes
primarily from the SB2 contribution, derived directly from the observed data, and
so is unaffected by uncertainties in the SB1 distribution. Nonetheless, the $q$
distribution is qualitatively similar if a zero offset is used in
equation~\ref{eqn:SB1_offset} instead of 0.2; quantitatively, the first four SB1
bins in the best fit with zero offset have frequencies of 32.3 instead of 28.3,
and all the remaining bins have frequencies of 16.1. This has the effect, if
anything, of accentuating the $q=1$ peak in the overall distribution. Curiously,
the zero offset case (which corresponds to the fainter component of the SB1 making
no significant contribution to the total luminosity) puts some stars into the
$q=0.9-1$ bin. These systems presumably have evolved primaries that are much
brighter than their unevolved companions.

The distribution with zero offset gives a hint of a second peak for low $q$, so it
is worth looking at the best-fitting $q$ distributions with a peak. These are
shown in Fig.\ref{fig:qSB2and1-pks}. However, the fit to the observed $f(m)/m_1$
distribution is much less good than for our preferred stepped $q$ distribution,
the peak for the best fit peaked distribution is not very pronounced (the best fit 
histogram is not very different from the one for a stepped distribution in 
Fig.\ref{fig:qSB2and1}), and the peak at $q=1$ is still dominant.

It is also interesting to compare our best-fitting $q$ distribution with the one
that would be predicted if we took the components at random from a steep IMF. This
prediction has been made by Tout (1991), who considered an IMF steep above
1\,M$_\odot$ but flat for smaller masses. For the SB2 distribution, he took the
lower mass cut-off for both components to be 1\,M$_\odot$ and found a curve that
rose steeply from $q=0$ to $q=1$, similar to our SB2 distribution, although not so
concentrated around $q=1$ . For the SB1 distribution he took the same lower mass
cut-off for the primary but chose 0.2\,M$_\odot$ as the cut-off for the secondary.
This gave a $q$ distribution with a strong peak at $q=0.2$ and a curve that
dropped smoothly to a low value at $q=1$ (see Figure 6 of Tout 1991). The joint
distribution is thus bimodal, similar to the result found by Staniucha (1979)
illustrated in Figure 1 of Tout (1991). This strongly double-peaked distribution
is not consistent with our $q$ distribution, which is very flat for $q<0.7$, even
for the peaked distribution that we tried (Figs~\ref{fig:qSB2and1} and
\ref{fig:qSB2and1-pks}). We conclude that the components in our sample of binaries
were {\em not} chosen independently and at random from the steep IMF that they 
seem to
obey (see discussion in the next section). A similar conclusion was reached by
Eggleton, Fitchett \&\ Tout (1989) for a more restricted sample of visual binaries
with two bright components.

Fig.\ref{fig:fracs_M_v} demonstrates that the fractions of SB2s and SB1s of all
stars (all entries in HIPPARCOS within 100\,pc and the same absolute magnitude
limit) behave rather similarly for varying limiting absolute magnitudes (the
deviation at brighter $M_{\rm{V}}$ for SB2s being due to low absolute numbers).
This to some degree justifies our using parameters derived from SB2s, such as the
value of $\alpha_0$, for SB1s as well. Note that here, as in the next section, we
are dealing with $M_{\rm{V}}$, and not $M_{\rm{V}_1}$, as it is the selection
biases on the \emph{systems} that we are interested in.

Given that the SB2s and SB1s seem to have essentially the same selection criteria,
the two distributions will also be independent of each other, the probability of
detecting one set of lines or two then depending only upon the detector
resolution. In this case, detecting one set of lines will be independent of
detecting the other and it is thus justifiable to add the two $q$ distributions
together in the way we have done to produce Fig.~\ref{fig:qSB2and1}.


        \section{Discussion}
        \label{sect:discussion}


So far, we have derived the observed period, primary mass and mass-ratio
distributions for our distance-limited sample. If there are any serious selection
effects, however, the true distributions could well be significantly different.
Selection effects acting on the $q$ distribution have already been discussed in
detail in Section~\ref{subsect:q distrib} and thus we can be confident of the
reality of the peak in the $q$ distribution near $q = 1$.

\begin{figure}
    \includegraphics[scale=0.35, angle=-90]{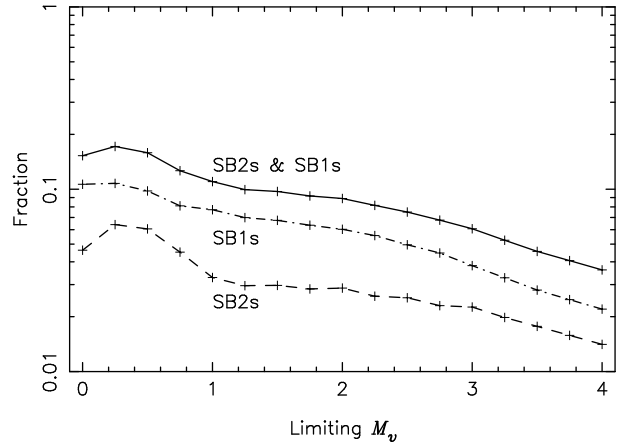} 
    \caption{Fraction of SBs, distance $\le 100$\,pc, for different
limiting $M_{\rm{V}}$. Fractions are defined as in Fig.~\ref{fig:Pdistribs}.}
    \label{fig:fracs_M_v}
\end{figure}

\begin{figure}
    \includegraphics[scale=0.35, angle=-90]{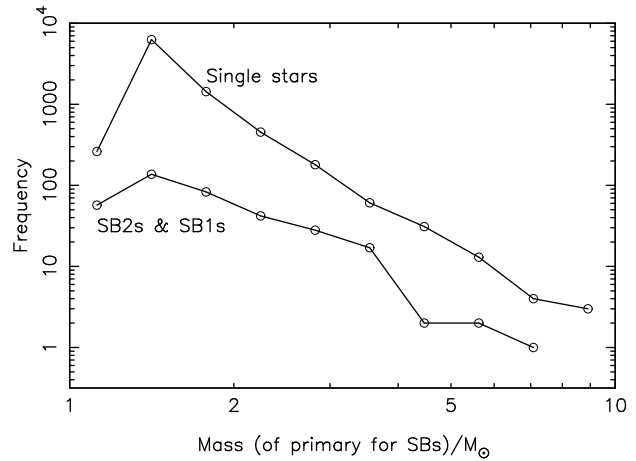} 
    \caption{Frequency distribution of masses (in solar masses) of
    SB primaries and of single stars.}
    \label{fig:m1_SBs_nonSBs}
\end{figure}

\begin{figure}
    \includegraphics[scale=0.35, angle=-90]{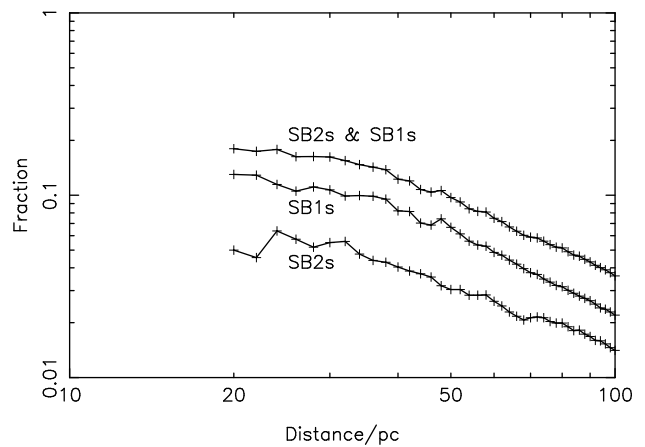} 
    \caption{Fraction of SBs, $M_{\rm{V}} \le 4$, for different volumes up
to 100\,pc. Fractions are defined as in Fig.~\ref{fig:Pdistribs}.}
    \label{fig:fracs_vol}
\end{figure}

There is however another obvious selection effect possibly acting on
 the $m_1$ distribution that has not yet been considered: the
 possibility of a lower detection rate for less luminous
 binaries. This would be reflected in a less pronounced increase in
 the observed $m_1$ distribution towards smaller masses compared to
 the true present-day mass function (PDMF). In the \emph{observed}
 distribution in Fig.~\ref{fig:m1_SBs_nonSBs}, $dN/d\log{m_1}
 \propto m_1^{-2.8}$, while for single stars it is approximately
 $\propto m^{-4.8}$ (Schr\"oder 1998). However, we need to know if
 this difference is genuine or due to a selection effect (or possibly
 both). To do this we look at the variation in detected SB fraction
 with volume (Fig.~\ref{fig:fracs_vol}) and compare it with the
 variation with limiting $M_{\rm{V}}$ within 100\,pc
 (Fig.~\ref{fig:fracs_M_v}).

\begin{table}
        \caption{Comparing the change in fraction with limiting
$M_{\rm V}$ (Fig. \ref{fig:fracs_M_v}) with the change with distance (Fig.
\ref{fig:fracs_vol}). 
}
    \begin{tabular}{p{3.7cm}p{3.7cm}}
        \hline
{\bfseries Fig. \ref{fig:fracs_M_v}:} change of fraction with $M_{\rm V}$. &
{\bfseries Fig. \ref{fig:fracs_vol}:} change of fraction with volume (max.
distance of sample).\\
        \hline
Decreasing average $m_{\rm V}$  \& decreasing average mass going from left to
right. & Decreasing average $m_{\rm V}$   but {\bfseries \emph{same}} average mass
going from left to right.\\
        \hline
Fraction falls by a factor of $\sim$ 4.3 from $M_{\rm V}$  = 1 to 4. & Fraction
falls by a factor of $\sim$ 5 from 25 to 100\,pc. \\
        \hline
Could be due to a shallower PDMF and/or a selection effect. & Could {\bfseries
\emph{only}} be due to an increasing incompleteness with increasing vol. \&
decreasing $m_{\rm V}$.\\
        \hline
    \end{tabular}
    \begin{tabular}{p{7.5cm}}
Factor is approximately the same, therefore the decrease with $M_{\rm{V}}$  (and
hence PDMF) is due to a selection effect. Therefore the PDMF of SB primaries is
the same as that of single field stars.\\
        \end{tabular}
        \label{table:selection_effects}
\end{table}
Table~\ref{table:selection_effects} summarizes the following argument. The
variation with $M_{\rm{V}}$ shows fractions with decreasing average mass and
decreasing average \emph{apparent} brightness as the absolute magnitude limit
becomes fainter, while the variation with volume shows fractions which again have
decreasing apparent brightness as the volume is increased but now have the
\emph{same} average mass for all volumes. The variation with volume shows a
decrease in fraction by a factor of $\sim 5$ from  25 to 100\,pc, while the
variation with limiting $M_{\rm{V}}$ shows a decrease by a factor of $\sim 4.3$
over a corresponding range of $M_{\rm{V}} = 1$ to 4 (a factor of 4 in distance
being equivalent to a difference of 3 in magnitude).  The decrease with limiting
$M_{\rm{V}}$ could again be due to a shallower PDMF or a selection effect, but the
decrease with volume could only be due to the increasing incompleteness as the
volume enlarges (and apparent brightness decreases). The fact that the two
fractions fall off by approximately the same factor shows that the decrease with
$M_{\rm{V}}$, and hence the shallower PDMF, is indeed due to a selection effect.
The true PDMF and IMF of the binary primaries are therefore nearly identical to
those of single field stars in the solar neighbourhood (Schr\"oder \& Pagel 2003
and references contained therein).


        \section*{Acknowledgments}


We wish to express our special gratitude to R.F. Griffin for the generous use of
his unpublished SB data which constitutes a significant fraction of the total
sample studied. We also thank the \emph{Centre de Donn\'ees astronomiques de
Strasbourg} (CDS) for their excellent internet database through which access was
gained to the \emph{Hipparcos} and Batten catalogues, and also the referee, Chris
Tout, and colleagues at the Astronomy Centre at the University of Sussex, who gave
much valuable advice on improving the paper. JF wishes to acknowledge the support
of a Postgraduate Assistantship from the University of Sussex.




\label{lastpage}
\end{document}